\documentclass[useAMS,usenatbib,usegraphicx]{mn2e}

\usepackage[latin1]{inputenc}
\usepackage{latexsym}
\usepackage{amsmath}

\newcommand{\unit}[1]{\ensuremath{\:\mathrm{#1}}}
\newcommand{\ie}{\mbox{i.e.}}

\title[Rotation speed and stellar axis inclination from p modes]{Rotation speed and stellar axis inclination from p modes:\\ How CoRoT would see other suns}
\author[J. Ballot, R. A. Garc\'\i a and P. Lambert]%
{J. Ballot,$^{1}$\thanks{\emph{Present address:} Max-Planck-Institut f{\"{u}}r Astrophysik, Karl-Schwarzschild-Str. 1, 85748 Garching, Germany; E-mail: {jballot@mpa-garching.mpg.de}}
R. A. Garc\'\i a$^{2,1}$
and P. Lambert$^{2,1}$\\
$^{1}$AIM -- Unit\'e mixte de recherche CEA-CNRS-Universit\'e Paris VII -- 
   UMR n$^o$ 7158, CEA/Saclay, 91191 Gif-sur-Yvette {\sc cedex}, France \\
$^{2}$DSM/DAPNIA/Service d'Astrophysique, CEA/Saclay,
	 91191 Gif-sur-Yvette {\sc cedex}, France}
\begin{document}

\date{Accepted 2006 March 24. Received 2005 November 18}

\pagerange{\pageref{firstpage}--\pageref{lastpage}} \pubyear{2006}

\maketitle

\label{firstpage}

\begin{abstract}
   {In the context of future space-based asteroseismic missions, we
have studied the problem of extracting the rotation speed and
the rotation-axis inclination of solar-like stars from the expected data. We have focused 
on slow rotators (at most twice solar rotation speed), firstly because they constitute the most difficult 
case and secondly because some of the CoRoT main targets are expected to have slow rotation rates. 
Our study of the likelihood function has shown a correlation
between the estimates of inclination of the rotation axis $i$ and the rotational splitting 
$\delta\nu$ of the star. By using the parameters, $i$ and $\delta\nu^{\star}=\delta\nu\sin i$, we propose and discuss new fitting strategies.
 Monte Carlo simulations have shown that we can extract a mean splitting 
and the rotation-axis inclination down to solar rotation rates. However, at the solar rotation rate we are not able to correctly recover the angle $i$ although we are still able to measure a correct $\delta\nu^{\star}$ with a dispersion less than 40 nHz. }   
\end{abstract}

\begin{keywords}
Rotation -- Stars: oscillations --
Sun: helioseismology -- Methods: data analysis -- Instrument: CoRoT
\end{keywords}  

%

\section{Introduction}
Understanding dynamical phenomena inside stars is 
one of the most important current challenges for stellar physics.
During the last decades, helioseismology has
allowed astrophysicists to constrain the internal structure and dynamics
of the Sun. In the same way, asteroseismology will aim to 
improve our knowledge of stellar dynamics, especially 
 convection and rotation.
With future asteroseismic missions like 
CoRoT
\citep[Convection Rotation and planetary Transits,][]{COROT03},
it will be possible for example to determine
the extent of the convective region in stars 
and to extract information on rotation.
Since 2003 the first Canadian satellite dedicated to asteroseismology,
MOST \citep[Microvariability and Oscillations of STars,][]{WalkerM03MOST},
has been operational, beginning the space age for asteroseismology.

Asteroseismology has already provided information on the internal rotation
of stars \citep[e.g.][ for results on a $\beta$ Cepheid]{AertsT03}.
However, the most accurate seismic information has
been obtained for the Sun.
Helioseismology has provided very accurate profiles
of the internal rotation
\citep[see][ and the references therein]{Thompson03ARA&A}
as deep as 0.2\:R$_{\sun}$ \citep{Couvidat03Rot,Garcia04Rot}, 
thanks to the Solar Heliospheric Observatory
and to ground-based networks.
Because of the rotation of stars, modes are not single peaks but multiplets.
The splitting of the multiplet components
gives information on the rotation speed in the acoustic cavity covered by
the mode. Nowadays and in the near future, the asteroseismic observations will be limited
to low-degree modes because of the absence of spatial resolution
on the stellar surface.
Thus, new inversion techniques have been developed and checked
to derive, for example, the radial rotation profile 
(e.g. \citeauthor{GoupilD96} \citeyear{GoupilD96}; \citeauthor*{LochardS04} \citeyear{LochardS04})
or to infer the latitudinal
differential rotation \citep{GizonS04}.

Rotational splittings could be derived from the oscillation spectrum 
along with the other
mode parameters.
However, as we have learned from the solar case, 
the rotational splitting is harder to extract for low-degree modes because of the limited number of components in a multiplet.
Moreover, another difficulty appears in the stellar case: the angle of inclination
($i$) of the rotation axis, which determines the multiplet pattern, is generally unknown.

\citet{GizonS03} (hereafter GS03) have recently studied the simultaneous extraction of
the splitting and the angle $i$ from low-degree oscillation modes.
We propose here to follow up their analysis by studying the potential of multi-mode fitting 
for more critical situations and by proposing automatable procedures.
Our main objective is to determine the precision and limits 
in the determination of the  
rotation of solar-like stars from a mission like CoRoT. To do so, we have simulated
CoRoT-type observations (150-day long) of a Sun spinning at different speeds with different axis orientations.
We have considered rather realistic signal-to-noise ratios (S/N) and we have
focused on the particular situation of
slow rotators (less than twice solar rotation).

Our preliminary results have been outlined in \citet{Ballot04SoHO}.
The present paper fully develops this work. 
The layout of the rest of the paper is as follows.
In Sect.~\ref{Sec:Spect} we describe the main properties of modes
for a star under rotational effects.
In Sect.~\ref{Sec:Extract} we describe the techniques used to extract splittings and angle $i$ 
from several modes together.
In Sect.~\ref{Sec:Simul} we present the results of our method applied
to several example cases.
Finally, we discuss the fitting methods before concluding
in the last section.

\section{Oscillation spectrum of a spinning star}\label{Sec:Spect}

\subsection{Mode properties}\label{SSec:ModeProp}

Acoustic (p) modes in solar-like stars are excited by turbulent
convective motions. Oscillations are damped but permanently re-excited
\citep*{GoldreichM94}.
The oscillation power spectrum of such modes can be modelled as a
noisy Lorentzian profile. For a power spectrum classically computed
with the Fourier transform of a regularly-sampled time series, 
this noise is a multiplicative exponential.
A mode $(n,\ell,m)$ -- see below -- is also characterized 
by its frequency, its amplitude
and its FWHM.

In solar-like stars, the width $\Gamma$ of a p-mode depends
only on its frequency $\nu$. For the Sun, the function $\Gamma(\nu)$
shows a S-shape. There is a plateau in the range  2300--3200\unit{\mu Hz}
around a value of 1\unit{\mu Hz}.
At low frequency widths decrease rapidly and increase at high
frequency \citep[e.g.][]{Garcia04YaleAct}.

In the absence of rotation the frequency of a mode depends only on
its radial order $n$ and its degree $\ell$:
we denote it $\nu_{n\ell}$.
Modes are ($2\ell+1$)-times degenerate
among the azimuthal order $m$. 
This degeneracy is removed by breaking the spherical symmetry,
especially by rotation. The frequency of mode $(n,\ell,m)$
is expressed as $\nu_{n\ell m}=\nu_{n\ell}+\delta\nu_{n\ell m}$. 
The asymptotic first-order approximation, 
developed for a star
spinning as a solid body with an angular velocity $\Omega$, gives
$\delta\nu_{n\ell m}=-m\delta\nu$ with $\delta\nu=\Omega/2\pi$
\citep{Ledoux51}.
We call $\delta\nu$ \textit{rotational splitting}
(or simply \textit{splitting}).

For geometrical reasons, only low-degree modes have a sufficient
amplitude to be visible in an oscillation spectrum due to the integration
of the luminosity -- or the radial velocity -- on the full stellar disk.
Mode amplitudes also depend on their azimuthal order $m$.
Calculations are rather straightforward and can be found for example in GS03.
Assuming the equipartition of energy between the different
components of a multiplet $(n,\ell)$, their amplitudes can
be expressed as
\begin{equation}
 A_{n\ell m} = a_{\ell m}(i) V_\ell^2  \alpha_{n\ell}
             = a_{\ell m}(i) A_{n\ell}.
\end{equation}
In this expression, the factor
$V_{\ell}$
is the mode visibility. It depends on the limb-darkening function, i.e. on the atmospheric properties.
The visibility $V_\ell$ decreases strongly when $\ell$ increases:
for $\ell=1,..,5$, we have calculated $(V_\ell/V_0)^2=1.5,0.53,0.027,0.0039,0.00067$,
assuming an Eddington law for the limb-darkening function.
For this reason, we expect to measure only modes $\ell=0$, 1, 2 and probably
a few $\ell=3$.
The factor $a_{\ell m}(i)$
is the amplitude ratio of modes inside a multiplet. It is a purely
geometrical term, depending on $i$, the angle between
the line of sight and the rotation axis.
This is true under only one condition, that the
contribution of each stellar-surface element to
the total flux depends only on its distance to the disk centre. Even if it is not exactly true for velocity-fluctuation observations due to the rotation of the star \citep[e.g.][]{ThHenney}, this
assumption stays very good for luminosity observations.
The final factor of the mode amplitude 
$\alpha_{n\ell}\approx\alpha(\nu_{n\ell})$
depends mainly on the frequency and excitation mechanisms.
We note $A_{n\ell}=V_{\ell}^2\alpha_{n\ell}$.
This approach is valid for low rotation rates, when rotation can be
interpreted as perturbation.

Thus a mode $(n,\ell)$ is modelled by a multiplet parame\-trized by five parameters
(only three for $\ell=0$): 
the central frequency $\nu_{n\ell}$, the amplitude $A_{n\ell}$, the width $\Gamma_{n\ell}$
common to all the components, 
the splitting $\delta\nu$ and the angle $i$.

\subsection{Classification depending on $\delta\nu$}
\begin{figure}
\centering
\includegraphics[width=\hsize]{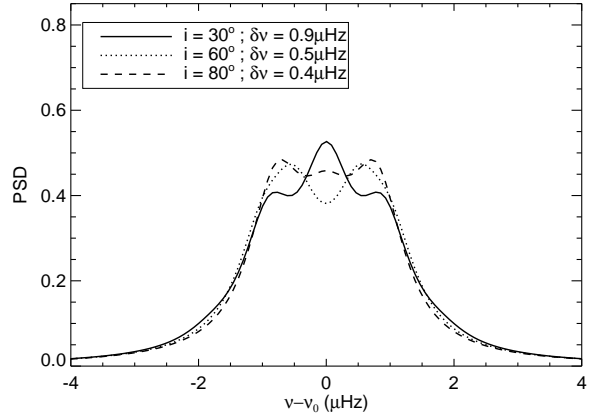}
\caption{An $\ell=2$ mode for three different speeds and angles.
\label{Fig:l2}}
\end{figure}

We have defined three different scenarios
according to 
$\delta\nu$:
\begin{enumerate}
\renewcommand{\theenumi}{(\arabic{enumi})}
\item $\delta\nu \gg \delta_{02}\nu$ ,
\item $\Gamma < \delta\nu \la \delta_{02}\nu$ ,
\item $\delta\nu \la \Gamma$ ;
\end{enumerate}
where $\delta_{02}\nu$ denotes the small separation 
$\nu_{n+1,\ell=0}-\nu_{n,\ell=2}$ 
(around 10\unit{\mu Hz} for the Sun in the range 2000--3000\unit{\mu Hz}).
In the first case, the components of different modes are mixed and
it could be difficult to label each peak in a spectrum
with the correct values of $\ell$, $m$ and relative $n$.
However, when this identification is done, all of the splittings
$\delta\nu_{n\ell m}$ are accurately defined.
In the second situation, 
mode identification does not pose any
problem in general for good S/N and, as the components of a multiplet are 
well separated,
splittings are easily measured.
In the third and last case, the multiplet components are blended. The effect
on the amplitude ratio of a multiplet due to a given inclination axis is not always distinguishable from
those of the splitting as illustrated by Fig.~\ref{Fig:l2}.
For three different configurations chosen as an example,
the mode profiles are nearly the same; only fine differences appear
in the structure of profile tops.
When an exponential multiplicative
noise is taken into account, these differences are very difficult to catch.
We have studied this more challenging situation, corresponding
to $\delta\nu \la 1\:\mathrm{\mu Hz}$ (for suns), i.e.
$\Omega \la 2\Omega_{\sun}$ 
($\Omega_{\sun}/2\pi\approx 0.4\:\mathrm{\mu Hz}$).

\section{Extracting the mode parameters}\label{Sec:Extract}

\subsection{Fitting modes: maximum likelihood}
Splittings and inclination angle should be deduced from the oscillation spectra
at the same time as all the other mode parameters. For that,
we use techniques developed and
applied in full-disk-integrated helioseismology.
Oscillation spectra are fitted with a maximum likelihood method as described by
\citet*{Appourchaux98a}. The power spectrum of a solar-like star is modelled as the sum of modes, modelled by multiplets, and a background noise, mainly due to convective motions (granulation, supergranulation), and instrumental noises.
The first step of the analysis is to remove the background, previously fitted
following the model of \citet{Harvey85}, to obtain a ``flat'' background.
Then the modes are classically fitted alone or by pairs
$(\ell,\ell+2)$ according to the value of the small separation and the
mode amplitudes. The residual background is considered as a constant
inside the fitting window.
As $i$ is a new parameter compared to the classic
helioseismic analysis,
we have explored its impact, especially on the
splitting determination.

\subsection{Guessing and assumptions}\label{SSec:guess}

The fitting method needs guesses for the parameters to fit.
This estimate is a starting point of the parameter-space exploration
by the algorithm maximising the likelihood.
We denote by $\tilde{x}$ the estimate of the parameter $x$.
A crude estimation of the mode central frequency can be 
obtained, by looking for
its centroid. The amplitudes and widths
can be first determined on
$\ell=0$ modes, which are insensitive to rotation.
As amplitudes $\alpha_{n\ell}$ and widths
$\Gamma_{n\ell}$ depend mainly on frequency,
initial values for the modes $\ell \ge 1$  can be interpolated 
from those of $\ell=0$ as follows:

\begin{align}
&\tilde{A}_{n-1,2}=\frac{V_2{}^2}{V_0{}^2}\tilde{A}_{n,0},
&
\tilde{A}_{n-1,1}=\frac{V_1{}^2}{V_0{}^2}\frac{\tilde{A}_{n-1,0}+\tilde{A}_{n,0}}{2},\\
&\tilde{\Gamma}_{n-1,2}=\tilde{\Gamma}_{n,0},
&
\tilde{\Gamma}_{n-1,2}=\frac{\tilde{\Gamma}_{n-1,0}+\tilde{\Gamma}_{n,0}}{2}.
\end{align}

Determining the estimates $\tilde{\imath}$ and $\tilde{\delta}\nu$ is not
easy when multiplet components are not well separated.
A first possibility is to fit each mode as a single Lorentzian.
The comparison between the widths of two neighbouring
modes $\ell=2$ and $\ell=0$ allows us to detect the presence of rotation (when $i > 0$),
but a quantitative interpretation of this broadening 
is difficult because of the cumulative effects of $i$ and $\delta\nu$.

The sensitivity of the fitting to $\tilde{\imath}$ and $\tilde{\delta}\nu$ has been tested along with
the impact of the noise on $i$ and ${\delta}\nu$ determination.
Modes of interest have been fitted as follows:
\begin{itemize}
\item
Pairs $\ell=0$ \&\ 2 are fitted with eight parameters
$(A_0,A_2,\nu_0,\nu_2,\Gamma,b,\delta\nu,i)$: their amplitudes, their
frequencies, a common width, the background level, the splitting and
the inclination angle.
Assuming $\Gamma_{n-1,2}=\Gamma_{n,0}$ is a good approximation as shown by the solar case \citep{ChaplinFLAG06}. Thus, the parameter space is reduced as well as the computing time and the risk of non convergence.
\item
Modes $\ell = 1$  are fitted with six parameters
$(A_1,\nu_1,\Gamma_1,b,\delta\nu,i)$. As the expected amplitudes of the $\ell=3$ modes are very small we do not fit them, although they are present in simulated spectra. 
{Previous results have shown that such a simplification could introduce biases -- especially to frequencies and splittings -- if neglected modes are not sufficiently small and/or are too close to the fitted ones. We have been careful and we have verified that no significant bias has been introduced in our case.}
\end{itemize}

\begin{figure}
\centering
\includegraphics[width=\hsize]{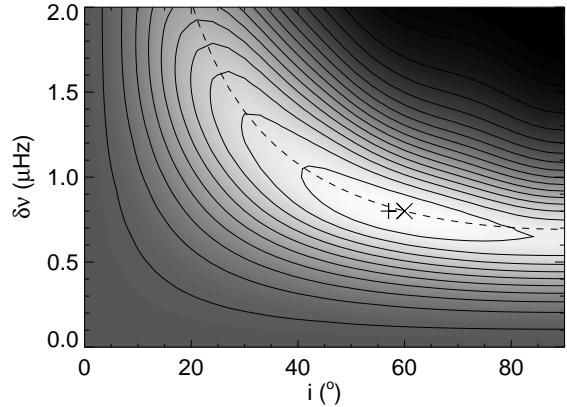}
\caption{Likelihood function for one simulated spectrum in the plane 
$(i,\delta\nu)$ of the parameter space. All the other parameters are 
fixed to their simulated value. The power spectrum is taken in the range 
2200--3000\unit{\mu Hz}.
The white colour corresponds to the highest likelihoods and the black 
to the lowest. The $\times$ is the simulated value $(i_0,\delta\nu_0)$ 
and the $+$ is the maximum of the likelihood. The dashed line
follows $\delta\nu\sin i = \delta\nu_0\sin i_0$.\label{Fig:dnusiniLH}}
\end{figure}

Different random values for $\tilde{\imath}$ and $\tilde{\delta}\nu$
have been tested on several Monte Carlo realizations of the spectrum
(see method in Sect.~\ref{SSec:MC}). We have seen first that
the solution is not unique and a certain dependence upon the first guess parameters is observed.
For a given mode in a given realization, the fitting procedure can converge
to some different couples $(i,\delta\nu)$ according to the initial
values $\tilde{\imath}$ and $\tilde{\delta}\nu$.
However the main effect is due to the noise which has a strong impact on the estimation of
$(i,\delta\nu)$ and disperses the results. Nevertheless, we observe a clear
correlation in the determination of both parameters. 
Results are organized along the curve:
$\delta\nu\sin i \approx \mathrm{constant} =\delta\nu_0\sin i_0$.
We denote with an index 0 $(\delta\nu_0,\sin i_0)$ the real (input)
values of the parameters in the simulation.
This can be explained by a study of the likelihood function
for a simulated spectrum. Figure~\ref{Fig:dnusiniLH} shows 
such function in a plane $(i,\delta\nu)$ in the parameter space,
with all other parameters fixed to their true value.
We observe in such a plane a ridge following the curve
 $\delta\nu\sin i = \delta\nu_0\sin i_0$.
Thus $i$ and $\delta\nu$ are correlated.
A new pair of independent parameters can be built:
\begin{equation}
(i,\delta\nu^{\star})\quad\mbox{ with }\delta\nu^{\star}=\delta\nu\sin i.
\end{equation}
Hereafter we use preferentially this new variable 
$\delta\nu^{\star}$ which is better suited than $\delta\nu$
for studying fitting issues and discussing results.
We did not find major differences between using
$\delta\nu^\star$ and $\delta\nu$ for the minimization routine, 
except for the error bars computed by Hessian-matrix inversion.

\subsection{Proposed strategy: multi-mode fitting}\label{SSec:Strategy}

With a classical fitting strategy, the determination of 
$i$ seems very sensitive and tricky.
We propose here another strategy aiming to improve the accuracy of
the obtained value of $i$.
We have fitted simultaneously several modes, as we can consider -- to a first order approximation -- that they have the same value of $i$ and $\delta\nu^\star$.
\begin{enumerate}
\renewcommand{\theenumi}{(\arabic{enumi})}
\item \textbf{Choosing initial guesses}.
For $\tilde{A}$, $\tilde\Gamma$ and $\tilde\nu$, see Sect.~\ref{SSec:guess}.
We firstly fit pairs $\ell = 0$ \&\ $2$ and single modes $\ell=1$, 
using several (typically 20) different random values of
$\tilde{\imath}$ and $\tilde{\delta}\nu^{\star}$.
We use the average of all the obtained results as a better guess for 
these parameters.
\item \textbf{Fitting simultaneously the modes
\boldmath$(\ell{=}2,n{-}1)$, $(\ell{=}0,n)$ and $(\ell{=}1,n)$\unboldmath}.
We used eleven parameters $(A_2, A_0, A_1, \nu_2, \nu_0, \nu_1, 
\Gamma_{0/2}, \Gamma_1, b, \delta\nu, i)$.
We obtained also a series of values for
$(i,\delta\nu^{\star})$. The mean values are noted
 $(i_{m}$ and $\delta\nu^{\star}_{m})$
($m$ for \textit{mean value}).
We obtain in this way a first measurement of $i$ and $\Omega$.
\item \textbf{Global fitting on a large range of the spectrum}. 
Fitting several modes simultaneously, keeping free all
the parameters, would be too costly in terms of computing time,
and too delicate in terms of convergence. So we have decided to
fix all the parameters but $(i,\delta\nu^{\star})$ to their values
deduced from the previous step.
We choose as guesses $\tilde{\imath}=i_{m}$ and 
$\tilde{\delta}\nu^{\star}=\delta\nu^{\star}_{m}$. 
Fitted results are denoted
$(i_{g},\delta\nu^{\star}_{g})$ ($g$ as \textit{global}).
\end{enumerate}

\section{Monte Carlo simulations}\label{Sec:Simul}
\subsection{Defining the simulations}\label{SSec:MC}

\begin{figure*}
\centering
\begin{tabular}{@{}c@{}c@{}}
a) $\Omega=2\Omega_{\sun}$ ($\delta\nu_0=0.8\unit{\mu Hz}$)&
b) $\Omega=\Omega_{\sun}$ ($\delta\nu_0=0.4\unit{\mu Hz}$)\\
\includegraphics[width=.5\hsize]{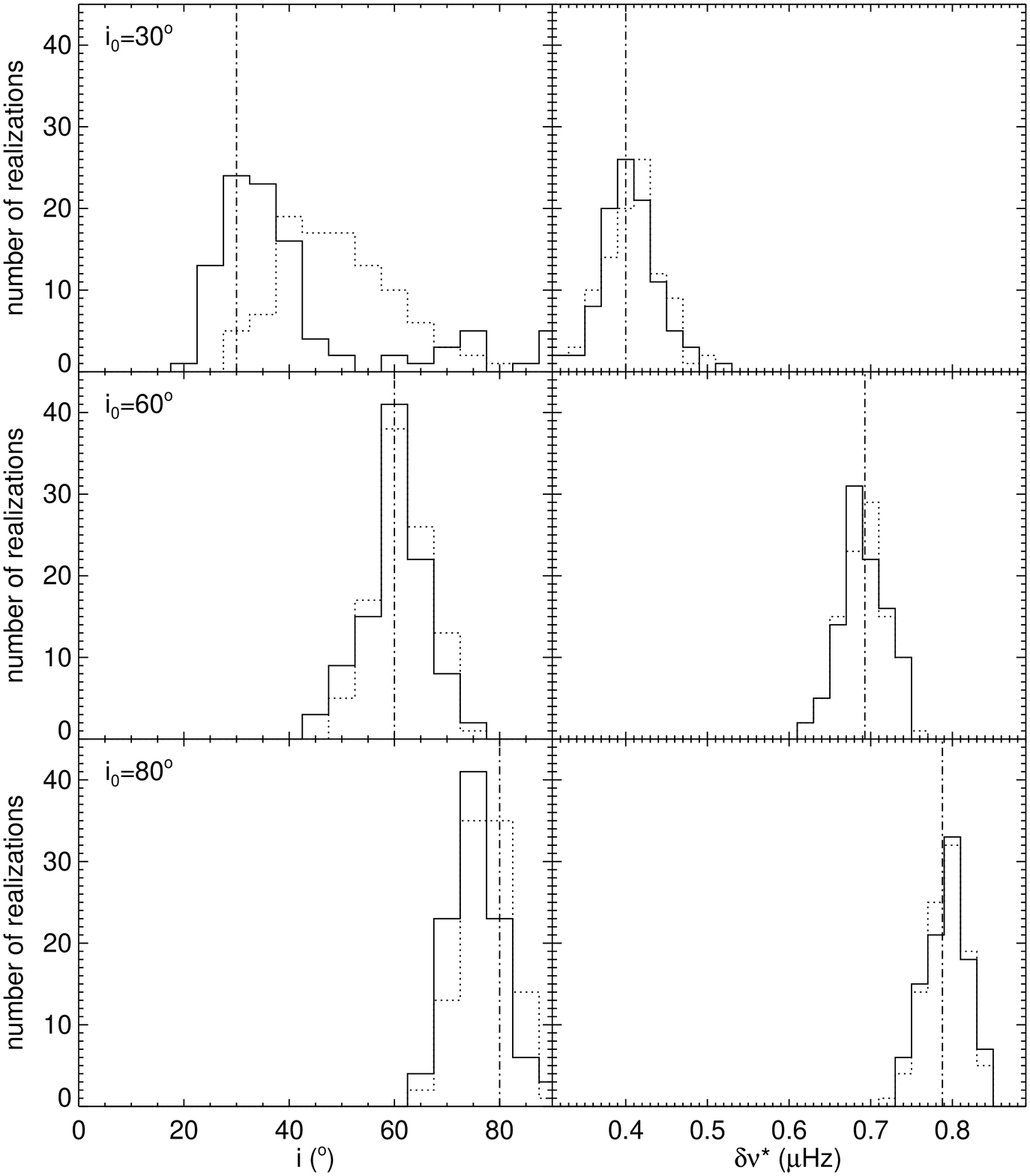}&%
\includegraphics[width=.5\hsize]{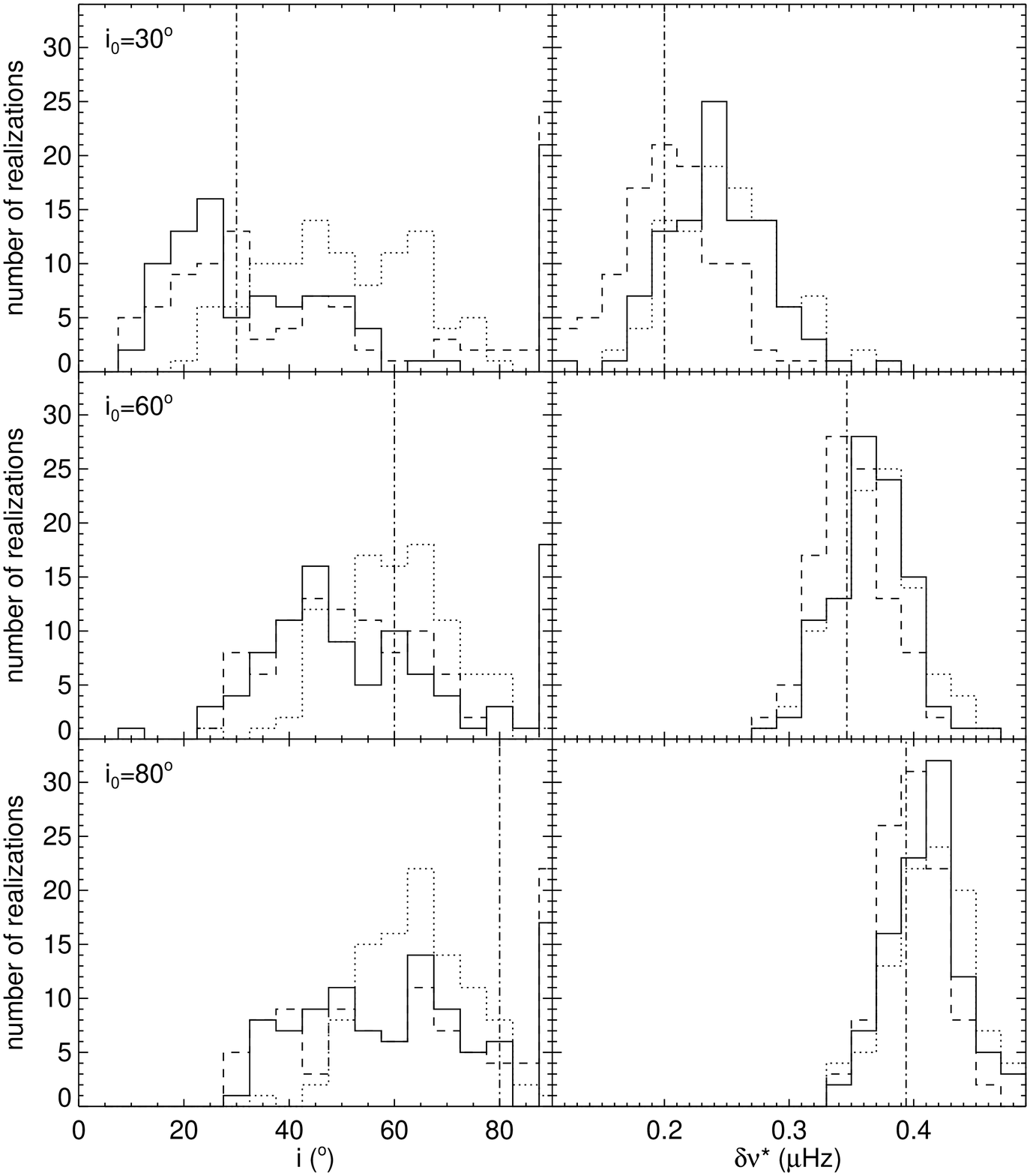}
\end{tabular}
\caption{a) Distribution of fitting results in the three configurations $\Omega=2\:\Omega_{\sun}$.
On the left the angle $i$; on the right $\delta\nu^{\star}$. Histograms
plotted with solid lines show the results of global fits
($i_g$ and $\delta\nu^{\star}_g$); Histograms with dotted lines show
the averaged results $i_m$ and $\delta\nu^{\star}_m$.
The dot-dash vertical line indicates the input value.
b) Same as a) but for  $\Omega=\Omega_{\sun}$.
We have added distribution of the ``idealized'' global fits (cf. text)
plotted with dashed lines.
\label{Fig:histoS0.8-0.4}}
\end{figure*}

The mode characteristics are derived from the observations of the Sun made by the GOLF instrument \citep[Global Oscillations at Low Frequency,][]{Gabriel95}. However the
amplitudes have been adapted to simulate luminosity observations instead of Doppler velocity measurements. 
We have treated the Sun as it was a main CoRoT target of magnitude ~6 observed
during 150 days.
For such a star, in the frequency range of interest (2200--3000\unit{\mu Hz}),
stellar noise have to dominate instrumental and photon noise \citep[see discussion in][]{Michel05}.
Thus, S/N \citep[as defined by][]{Libbrecht92} of the hightest
component of a multiplet varies from 15 to 150 for the $\ell=1$ modes,
from 4 to 45 for $\ell=2$, and from 0.7 to 7 for $\ell=3$ (in configurations at
80\degr). The widths do not vary much (from 0.8 to 1.1\unit{\mu Hz}).
150-day (resolution
$\approx 77\unit{nHz}$) power spectra are created including $\ell \le 3$  with a splitting
$\delta\nu_0$ and an angle $i_0$ that we want to simulate. 
In the chosen frequency range, there are six modes for each degree. 
This choice of interval results from a compromise: we have rejected modes with
too low S/N (i.e. at low frequency) and peaks too broad, useless for our analysis (at higher frequencies). This will give us a lower limit of what we could obtain in the real case, with the hope that CoRoT will reach
such modes. 
To introduce the noise of each realization we follow \citet{Fierry98} by using a random exponential distribution which simulates the stochastic excitation. 

To test the analysis method a Monte Carlo simulation is done, i.e., we repeat $N$ times the method on the same theoretical spectrum changing only the realization of noise. As the computing time required in each realization is quite high and we want to do many different cases, we have decided to limit the number of realizations $N$ to 100. The statistical significance of the results is small but it is enough to check the general trends of the solution. In order to verify our results we have increased $N$ to 1000 in some cases, e.g. $i_0=60$\degr, $\delta\nu_0=0.8\unit{\mu Hz}$. The conclusions remain roughly the same.
We have simulated six different configurations:
two rotation rates $\Omega = 1$ and $2\:\Omega_{\sun}$, \ie\ 
$\delta\nu_0 = 0.4$ and $0.8\unit{\mu Hz}$, with three inclination angles
$i_0$ = 30, 60 and 80\degr.

\subsection{Star spinning twice as faster as the Sun}

This class of stars is the most favourable among those considered.
Results obtained with our strategy are satisfactory.
A clear improvement is found, relative to classical fitting.
The histograms of Fig.~\ref{Fig:histoS0.8-0.4}-a show the distributions
of deduced parameters for each considered inclination angle.
Both parameter couples $(i_m,\delta\nu^{\star}_m)$ and
 $(i_g,\delta\nu^{\star}_g)$ are plotted for every studied stellar orientation.
We make three main comments:
\begin{itemize}
\item in the three configurations, 
determinations of $\delta\nu^{\star}$ are non-biased and little spread:
the dispersion is around 30\unit{nHz}.
Results given by averaging ($\delta\nu^{\star}_m$) and by global fitting
 ($\delta\nu^{\star}_g$) are very similar. Global fitting does not lead to
a noticeable change in this parameter in this situation.
\item On the other hand, the global fit (i.e. $i_g$) brings, for $i$, a major
improvement at low angle ($i=30$\degr) according to averaged results $i_m$.
Although there continue to be several highly spurious results 
($i_g \ga 70$\degr), a large number of realizations lie around 
 30\degr.

\item There is a slight bias on the $i$ determination for 
the extreme values, but it remains smaller than the error bar.
\end{itemize}

\subsection{Star spinning as the Sun}

Fitting results for the configuration with
$\delta\nu=0.4\unit{\mu Hz}$ are shown
in Fig.~\ref{Fig:histoS0.8-0.4}-b. The study of the distributions of 
$i_m$, $\delta\nu^{\star}_m$, $i_g$ and $\delta\nu^{\star}_g$
leads to two different conclusions for
$\delta\nu^{\star}$ and $i$.
\begin{itemize}
\item The $\delta\nu^{\star}$ distributions are quite narrow
with dispersions similar to the previous configurations (around
30--40\:nHz). However a significant bias appears in the three
cases, whereas it is negligible in the simulations at
$2\:\Omega_{\sun}$.
\item The angle $i$ is not correctly extracted.
The distributions are rather chaotic. 
However we have remarked that around a fifth of the realizations
have given an angle of 90\degr.
For the global fits of these low-splitting cases, this value 
behaves like an attractor during the likelihood-maximising process.
\end{itemize}

We wanted to know if it is possible to extract the angle $i$
from the selected modes in a configuration $\Omega=\Omega_{\sun}$.
To do so we have considered an idealized situation:
we have performed ``ideal'' global fits. In such fits all the
parameters -- except $i$ and $\delta\nu^{\star}$ -- 
are fixed to their exact values  and not to
the values deduced from a previous fitting step (cf. step \#3 in the
strategy \S~\ref{SSec:Strategy}). Moreover the exact values
$i_0$ and $\delta\nu_0^{\star}$ are chosen as guesses
$\tilde{\imath}$ and $\tilde{\delta}\nu^{\star}$.
Thus all is optimized for fitting: only noise can influence the
results. Results of this fitting method are plotted
in Fig.~\ref{Fig:histoS0.8-0.4}-b with dashed lines.
Thus we can conclude that:
\begin{itemize}
\item the bias on $\delta\nu^{\star}$ disappears. It indicates that
this bias was due to 
errors in the values to which the parameters were fixed.
However, the dispersion stays the same: it is mainly
generated by the noise.
\item the determination of $i$ is not changed. Noise dominates above
the signature of the angle and that seems inevitable in such data.
\end{itemize}

\section{Discussion}\label{Sec:Discuss}

\subsection{On the $i$/$\delta\nu$-correlation and the law $a_{\ell m}(i)$}
In the framework of global helioseismology,
\citet{ChaplinE01} have observed that changing amplitude ratios
fixed inside the multiplets $\ell=1,2$ and $3$ during the fit
of solar spectra introduces a systematic bias on extracted splitting. 
We can understand the reason by studying the likelihood function 
shown in Fig.~\ref{Fig:dnusiniLH}. 
Changing the amplitude ratio is similar to changing the angle $i$, thus
it introduces a bias on splitting determination
due to the correlation we have found.
Our results generalise this observation.
They show perfectly that we should be cautious of 
bias introduced when parameters are fixed, because of
the correlation existing between the different parameters
\citep[see also][]{Fierry98}.

This analysis shows that it could be interesting to derive
the angle $i$ by other ways, like directly studying the
light curve of stars and trying to follow up modulations due to 
activity spots \citep[e.g.][]{Ricinski04}.
If such an additional constraint is available, the situation would
become similar to the solar case and the amplitude ratios $a_{\ell m}$
could be fixed \textit{a priori} and individual splittings fitted.
However the measurement of $i$ must be sufficiently accurate
(probably $\sim5$--8\degr) otherwise the  estimate of $\delta\nu$ will likely be 
biased.

The results presented in this paper depend on the law we have
used to link $a_{\ell m}$ to $i$. For fitting, this law must
be defined \textit{a priori}. As shown once again by solar experience
\citep{FLAG04Yale}, when multiplet components are blended and not
separated -- which is the case here -- 
fits are very sensitive to the chosen law $a_{\ell m}(i)$.
Luckily, for intensity observations, these ratios depend
mainly on well-controlled geometrical considerations
(cf. Sect.~\ref{SSec:ModeProp}).

\subsection{Limitation and improvements}

The situation can be improved if low-frequency modes are measured. 
For these modes the splitting can directly be measured
because of their finer widths. Then fixing the retrieved splitting 
can yield to a good estimation of $i$ at higher frequencies 
where the multiplets are better defined and the influence 
of the stochastic excitation less important.

In our simulations, we have assumed that the angle and the splitting 
are the same for all modes.
While it is true that $i$ is the same for every mode, $\delta\nu$ can
vary for real stars, especially because of the differential
rotation that could exist along the radius. 
However, for the Sun this variation is weak for
low-degree modes in the studied frequency range. We could also
attempt to extract not a mean splitting but a mean splitting for
each degree, as was done for a first stage for the Sun
\citep[cf.][]{LazrekP96}.

If $\ell=3$ modes have sufficiently high amplitudes in real observations to be
correctly fitted, the results shown here would be improved.
If they could be observed but with low S/N, we could try to
use
a so-called $n$-collapsogram \citep[cf.][]{Ballot04SoHO}
to extract a mean splitting.
This technique can be summarized as follows:
averaging the spectra of several $\ell=3$ modes with different
orders $n$, after removing the $\ell=1$ neighbours, to enhance
the S/N and define the multiplet better; and fitting the resulting
spectrum.
It needs a good determination of the central frequency for every
mode, and small variations of $\delta\nu$ and the width
(which is the case in the ``plateau'' frequency range). 

We can hope to derive even better results by denoising
asteroseismic spectra.
Filtering the spectrum and enhancing S/N could improve the contrast of
multiplets, guiding their analysis. \citet{LambertP05}
are proposing methods based on curvelet transforms permitting
such denoising.

This analysis performed on the Sun can be extended to sufficiently bright 
CoRoT targets with similar mass and evolutionary state. S/N will depend on the 
convective-noise level which will be observed in other stars. 
Some discussions on this topic have taken place after the observations of
Procyon by MOST \citep{Matthews04,Bedding05}.

\section{Conclusion}

\begin{figure}
\centering
\includegraphics[width=\hsize]{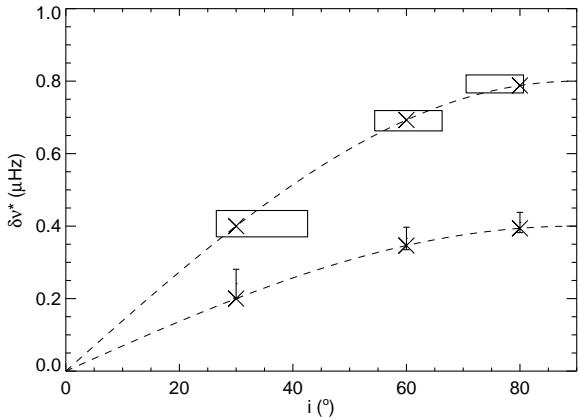}
\caption{Synthetic representation of biases and error bars for
$i$ and $\delta\nu^{\star}$  deduced from the simulations, in
all the studied configurations.
The crosses ($\times$) mark the expected values 
$(i_0,\delta\nu_0^{\star})$. For $\Omega=2\unit{\Omega_{\sun}}$ cases,
the boxes indicate the mean results and their dispersions. For
$\Omega=1\unit{\Omega_{\sun}}$ cases, only error bars on $\delta\nu^{\star}$
are plotted because of the absence of good determinations of $i$.
The two dashed lines are isorotations
$\delta\nu=\delta\nu_0=0.4\mbox{ and }0.8\unit{\mu Hz}$.\label{Fig:resS0.8-0.4tout}}
\end{figure}

One of the challenges of present and future asteroseismic space missions is to 
extract stellar rotation rates and, wherever possible, the internal rotation profile. 
To do that, mode splittings $\delta\nu$ have to be measured.
We have studied the impact of the extra parameter $i$, appearing in asteroseismology, on the fitting.
We have shown a correlation between $\delta\nu$ and $i$, and defined a new parameter
$\delta\nu^{\star}=\delta\nu\sin i$. Strategies of multi-mode fitting
have been developed, tested and validated with Monte Carlo simulations. Figure~\ref{Fig:resS0.8-0.4tout} sums up the results.
In agreement with GS03, we find that at $\Omega=2\unit{\Omega_{\sun}}$ 
we can retrieve both parameters in most of the cases, 
but with error bars improved  by the global fitting, especially at low angle.
However, at $\Omega={\Omega_{\sun}}$ we have not been able to correctly recover the angle $i$. This result emphasizes the interest of having an independent measurement of the angle, but it has to be accurate enough to prevent the inclusion of a bias in the splitting determination.

\section*{Acknowledgments}
The authors want to thank S. Turck-Chi\`eze and C. Catala for
their useful remarks.

\bibliographystyle{mn2e}
\bibliography{publi_rotmnras}

\begin{thebibliography}{}

\bibitem[\protect\citeauthoryear{{Aerts}, {Thoul}, {Daszy{\' n}ska},
  {Scuflaire}, {Waelkens}, {Dupret}, {Niemczura} \& {Noels}}{{Aerts}
  et~al.}{2003}]{AertsT03}
{Aerts} C.,  {Thoul} A.,  {Daszy{\' n}ska} J.,  {Scuflaire} R.,  {Waelkens} C.,
   {Dupret} M.~A.,  {Niemczura} E.,    {Noels} A.,  2003, Sci., 300, 1926

\bibitem[\protect\citeauthoryear{{Appourchaux}, {Gizon} \&
  {Rabello-Soares}}{{Appourchaux} et~al.}{1998}]{Appourchaux98a}
{Appourchaux} T.,  {Gizon} L.,    {Rabello-Soares} M.-C.,  1998, \aaps, 132,
  107

\bibitem[\protect\citeauthoryear{{Baglin}}{{Baglin}}{2003}]{COROT03}
{Baglin} A.,  2003, Advances in Space Research, 31, 345

\bibitem[\protect\citeauthoryear{{Ballot}, {Garc{\'{\i}}a}, {Lambert} \&
  {Teste}}{{Ballot} et~al.}{2004}]{Ballot04SoHO}
{Ballot} J.,  {Garc{\'{\i}}a} R.~A.,  {Lambert} P.,    {Teste} A.,  2004, in
  {Danesy} D.,  ed., ESA SP-559, SOHO14/GONG2004 Workshop, New
  Haven, p.~309

\bibitem[\protect\citeauthoryear{{Bedding} et~al.}{{Bedding} et~al.}{2005}]{Bedding05}
{Bedding} T.~R. et al., 2005, \aap, 432, L43

  
\bibitem[\protect\citeauthoryear{{Chaplin}, {Appourchaux}, {Baudin} \& {et
  al.}}{{Chaplin} et~al.}{2004}]{FLAG04Yale}
{Chaplin} W.~J. {et al.}, 2004, in {Danesy}
  D.,  ed., ESA SP-559, SOHO14/GONG2004 Workshop, New Haven,
  p.~356

\bibitem[\protect\citeauthoryear{{Chaplin}, {Elsworth}, {Isaak}, {Marchenkov},
  {Miller} \& {New}}{{Chaplin} et~al.}{2001}]{ChaplinE01}
{Chaplin} W.~J.,  {Elsworth} Y.,  {Isaak} G.~R.,  {Marchenkov} K.~I.,  {Miller}
  B.~A.,    {New} R.,  2001, \mnras, 327, 1127

\bibitem[\protect\citeauthoryear{{Chaplin} et~al.}{{Chaplin} et~al.}{2006}]{ChaplinFLAG06}
{Chaplin} W.~J. et al.,  2006, \mnras, accepted

\bibitem[\protect\citeauthoryear{{Couvidat}, {Garc{\'{\i}}a}, {Turck-Chi{\`
  e}ze}, {Corbard}, {Henney} \& {Jim{\' e}nez-Reyes}}{{Couvidat}
  et~al.}{2003}]{Couvidat03Rot}
{Couvidat} S.,  {Garc{\'{\i}}a} R.~A.,  {Turck-Chi{\` e}ze} S.,  {Corbard} T.,
  {Henney} C.~J.,    {Jim{\' e}nez-Reyes} S.,  2003, \apjl, 597, L77

\bibitem[\protect\citeauthoryear{{Fierry Fraillon}, {Gelly}, {Schmider},
  {Hill}, {Fossat} \& {Pantel}}{{Fierry Fraillon} et~al.}{1998}]{Fierry98}
{Fierry Fraillon} D.,  {Gelly} B.,  {Schmider} F.~X.,  {Hill} F.,  {Fossat} E.,
     {Pantel} A.,  1998, \aap, 333, 362

\bibitem[\protect\citeauthoryear{{Gabriel}, {Grec}, {Charra} \& {et
  al.}}{{Gabriel} et~al.}{1995}]{Gabriel95}
{Gabriel} A.~H. {et al.}, 1995, \solphys, 162, 61

\bibitem[\protect\citeauthoryear{{Garc{\'{\i}}a}, {Corbard}, {Chaplin} \& {et
  al.}}{{Garc{\'{\i}}a} et~al.}{2004a}]{Garcia04Rot}
{Garc{\'{\i}}a} R.~A. {et al.}, 2004a,
  \solphys, 220, 269

\bibitem[\protect\citeauthoryear{{Garc{\'{\i}}a}, {Jim\'enez-Reyes},
  {Turck-Chi{\` e}ze}, {Ballot} \& {Henney}}{{Garc{\'{\i}}a}
  et~al.}{2004b}]{Garcia04YaleAct}
{Garc{\'{\i}}a} R.~A.,  {Jim\'enez-Reyes} S.~J.,  {Turck-Chi{\` e}ze} S.,
  {Ballot} J.,    {Henney} C.~J.,  2004b, in {Danesy} D.,  ed., ESA SP-559,
  SOHO14/GONG2004 Workshop, New Haven, p.~436

\bibitem[\protect\citeauthoryear{{Gizon} \& {Solanki}}{{Gizon} \&
  {Solanki}}{2003}]{GizonS03}
{Gizon} L.,  {Solanki} S.~K.,  2003, \apj, 589, 1009

\bibitem[\protect\citeauthoryear{{Gizon} \& {Solanki}}{{Gizon} \&
  {Solanki}}{2004}]{GizonS04}
{Gizon} L.,  {Solanki} S.~K.,  2004, \solphys, 220, 169

\bibitem[\protect\citeauthoryear{{Goldreich}, {Murray} \& {Kumar}}{{Goldreich}
  et~al.}{1994}]{GoldreichM94}
{Goldreich} P.,  {Murray} N.,    {Kumar} P.,  1994, \apj, 424, 466

\bibitem[\protect\citeauthoryear{{Goupil}, {Dziembowski}, {Goode} \&
  {Michel}}{{Goupil} et~al.}{1996}]{GoupilD96}
{Goupil} M.-J.,  {Dziembowski} W.~A.,  {Goode} P.~R.,    {Michel} E.,  1996,
  \aap, 305, 487

\bibitem[\protect\citeauthoryear{{Harvey}}{{Harvey}}{1985}]{Harvey85}
{Harvey} J.,  1985, in {Rolfe} E.,  {Battrick} B.,  eds, ESA SP-235,
  Future Missions in Solar, Heliospheric \& Space Plasma Physics Workshop,
  Garmisch-Partenkirschen, p.~199

\bibitem[\protect\citeauthoryear{{Henney}}{{Henney}}{1999}]{ThHenney}
{Henney} C.~J.,  1999, PhD thesis, Univ. California

\bibitem[\protect\citeauthoryear{{Lambert}, {Pires}, {Ballot}, {Garc{\'{\i}}a},
  {Starck} \& {Turck-Chi{\`{e}}ze}}{{Lambert} et~al.}{2006}]{LambertP05}
{Lambert} P.,  {Pires} S.,  {Ballot} J.,  {Garc{\'{\i}}a} R.~A.,  {Starck}
  J.-L.,    {Turck-Chi{\`{e}}ze} S., 2006, \aap, accepted

\bibitem[\protect\citeauthoryear{{Lazrek}, {Pantel}, {Fossat} \& {et
  al.}}{{Lazrek} et~al.}{1996}]{LazrekP96}
{Lazrek} M. {et al.}, 1996, \solphys, 166, 1

\bibitem[\protect\citeauthoryear{{Ledoux}}{{Ledoux}}{1951}]{Ledoux51}
{Ledoux} P.,  1951, \apj, 114, 373

\bibitem[\protect\citeauthoryear{{Libbrecht}}{{Libbrecht}}{1992}]{Libbrecht92}
{Libbrecht} K.~G.,  1992, \apj, 387, 712

\bibitem[\protect\citeauthoryear{{Lochard}, {Samadi} \& {Goupil}}{{Lochard}
  et~al.}{2004}]{LochardS04}
{Lochard} J.,  {Samadi} R.,    {Goupil} M.,  2004, \solphys, 220, 199

\bibitem[\protect\citeauthoryear{{Matthews}, {Kusching}, {Guenther}, {Walker}, 
 {Moffat}, {Rucinski}, {Sasselov} \& {Weiss}}{{Matthews}
  et~al.}{2004}]{Matthews04}
{Matthews} J.~M., {Kusching} R., {Guenther} D.~B., {Walker} G.~A.~H., 
{Moffat} A.~F.~J., {Rucinski} S.~M., {Sasselov} D., {Weiss} W.~W., 2004,
{\nat}, 430, 51

\bibitem[\protect\citeauthoryear{{Michel}, {Samadi}, {Baudin}, {Auvergne} \& {the Corot team}}{{Michel}
  et~al.}{2005}]{Michel05}
{Michel} E. et al., 2005, Mem.S.A.It., 75, 282


\bibitem[\protect\citeauthoryear{{Rucinski}, {Walker}, {Matthews} \& {et
  al.}}{{Rucinski} et~al.}{2004}]{Ricinski04}
{Rucinski} S.~M. {et al.}, 2004,
  \pasp, 116, 1093

\bibitem[\protect\citeauthoryear{{Thompson}, {Christensen-Dalsgaard}, {Miesch}
  \& {Toomre}}{{Thompson} et~al.}{2003}]{Thompson03ARA&A}
{Thompson} M.~J.,  {Christensen-Dalsgaard} J.,  {Miesch} M.~S.,    {Toomre} J.,
   2003, \araa, 41, 599

\bibitem[\protect\citeauthoryear{{Walker}, {Matthews}, {Kuschnig} \& {et
  al.}}{{Walker} et~al.}{2003}]{WalkerM03MOST}
{Walker} G.  {et al.}, 2003, \pasp, 115, 1023

\end{thebibliography}
\label{lastpage}

\end{document}